\documentclass[a4paper]{article}
\usepackage{amsmath,amsthm,amssymb}
\usepackage{mathtext}
\usepackage{graphicx}
\usepackage{enumerate}
\usepackage{geometry}
\usepackage{url}
\usepackage{hyperref}
\usepackage{floatrow}
\usepackage{layout}
\usepackage{amssymb} 
\usepackage{multirow}
\usepackage{caption}
\usepackage{authblk}
\usepackage{dsfont}
\usepackage{braket}
\usepackage{xcolor}
\usepackage{cite}
\usepackage{graphicx}
\usepackage{dcolumn}
\usepackage{bm}
\usepackage{indentfirst}

\newcommand{\mean}[1]{\langle {#1} \rangle}
\newcommand{\prjct}[1]{\mathinner{|{#1}\rangle}\!\!\mathinner{\langle{#1}|}}
\newcommand{\ii}{\mathrm{i}}
\newcommand{\tr}{\text{tr}}

\renewcommand{\t}[1]{\mathrm{#1}}
\newcommand{\be}{\begin{equation}}
\newcommand{\ee}{\end{equation}}

\begin{document}
\title{Long-distance quantum key distribution based on the physical loss control}

\author{N.\,S.\,Kirsanov, N.\,R.\,Kenbaev, A.\,B.\,Sagingalieva, D.\,A.\,Kronberg, V.\,M.\,Vinokur, G.\,B.\,Lesovik\\
Terra Quantum AG, St. Gallerstrasse 16A, CH-9400 Rorschach, Switzerland}
\maketitle
\begin{abstract}
Existing quantum cryptography is resistant against secrecy-breaking quantum computers but suffers fast decay of the signal at long distances. The various types of repeaters of propagating quantum states have been developed to meet the challenge, but the problem is far from being solved. We step in the breach and put forth long-distance high secrecy optical cryptography, creating the fast quantum key distribution over distances up to 40,000 kilometers. The key element of the proposed protocol is the physical control over the transmission line.
\end{abstract}
\section{Introduction}

Existing quantum cryptography is robust against secrecy-breaking quantum computers but is subject to the fast decay of the signal with the distances. To amend the signal losses, the various types of repeaters of the propagating quantum states have been developed\,\cite{Kimble_Qrepeater, Sangouard_Qrepeater, Simon_Qrepeater, Duan_Qrepeater, Briegel_Qrepeater, Kok_Qrepeater, Childress_Qrepeater, van_Loock_Qrepeater, Wang_Qrepeater, Sangouard2_Qrepeater, Azuma_Qrepeater, Zwerger_Qrepeater, Munro_Qrepeater, Jiang_Qrepeater, Munro2_Qrepeater, Grudka_Qrepeater}, but yet the long distance transmission looks like an impossible task for existing protocols. 
Directly applying the principle of quantum irreversibility\,\cite{Lesovik_Qirreversibility, Kirsanov_Qirreversibility, Lesovik2_Qirreversibility}, we step in the breach and put forth long-distance high secrecy cryptography, creating the fast quantum key distribution over the globe distances of about 40,000 kilometers.


On a practical level,  the information transfer efficiency, both in classical and quantum cases, is hindered by the optical fiber losses. 
The conventional approach to quantum communication is double-suffering:
firstly, the losses themselves harm efficiency; 
secondly, the commonly accepted concept is that an eavesdropper (Eve) can use all the transmission line losses to decipher the communication successfully.
However, the lion share of losses occurs due to Rayleigh scattering of the signal propagating through the optical fiber. 
This propagation is similar, to no small extent, to the evolution of the ensemble of particles experiencing scattering on the quenched disorder potential described by the kinetic equation\,\cite{Landau_X}, generalizing the classical Boltzmann equation. 
This implies that the particle dynamics is accompanied by the entropy growth and, therefore, irreversible as expressed by the Second Law of Thermodynamics. 
Taking that the line carries a scatterer per a wavelength, we conclude that at least about $10^9$ quantum Maxwell Demon-like devices are required to collect photons scattered in a one kilometer long fiber and to reverse and unify the dynamics of the related quantum states. 
This makes collection of the Rayleigh-scattered information an impossible task.
These considerations supply us with a new paradigm for establishing an innovative quantum key distribution (QKD) protocol based on the physical control of the optical fiber line.






In a basic optical QKD scheme electromagnetic states representing different bit values are transmitted from the sender (Alice) to the receiver (Bob) which allows them to securely share a secret random sequence.
As we have just pointed out, the major irreversibility’s implication is that Eve cannot collect any useful information about the random bits from the scattering. 
However, the possibility of the local rerouting of the part of the transmitted signal remains. 
Physically this can be implemented, for example, by a local bending the fiber, which leads to mixing of the major propagating mode with the higher-order leaking modes. 
This local imperfection, in turn, allows for eavesdropping on the information carried by the major mode. 
Had the signal been classical, this “bending” would have opened unlimited access to the full content of the message for Eve. 
However, any electromagnetic signal is quantized and can be viewed as a sequence of photons. 
The discreet statistics of photons imposes a major limitation on Eve's ability to extract information from measuring seized signal.
If the initial signal contains, on average, $N$ photons, and the local leakage is quantified by transparency $r_E$, only the small fraction of the signal containing $n_E=r_EN$ photons comes to Eve. 
In the case of the coherent signal state, the  unavoidable fluctuation of the photon number follows the Poisson statistics, $\delta n_E=\sqrt{r_EN}$.
Therefore, with the decrease of $r_E$, the relative fluctuations grow as $\delta n_E/n_E=1/\sqrt{r_EN}$.
If, for instance, Alice encodes bit values into pulses with two different intensities, high fluctuations make it difficult for Eve to distinguish them.
In turn, if Bob gets the signal pulse comprising $N_{B}$ photons for which fluctuations $\delta N_{B}\ll N_{B}$, then even a single-shot measurement provides him with a high-probability reliable recognition of whether the received signal represents 0 or 1.
This establishes that the critical condition of security guarantee is ensuring that Bob gets the share sufficient for identifying the recognition of the signal, while the controlled value of $r_E$ limits the distinguishability of the signal portion seized by Eve.

An efficient control over $r_E$ rests on a careful analysis of the optical fiber's state and the emergent scattering matrix that can be accomplished by standard telecom technology methods, particularly, using the optical time-domain reflectometry.
We propose an efficient technique for controlling the optical fiber based on the direct measurement of the propagation of the signal from Alice to Bob.
This technique, together with the cascade signal amplification as a mean for preserving signal intensity lays the foundation for our QKD protocol unlocking unprecedented key generation rates and global transmission distances.



\begin{figure}[t]
    \noindent\centering{
    \includegraphics[width=\columnwidth]{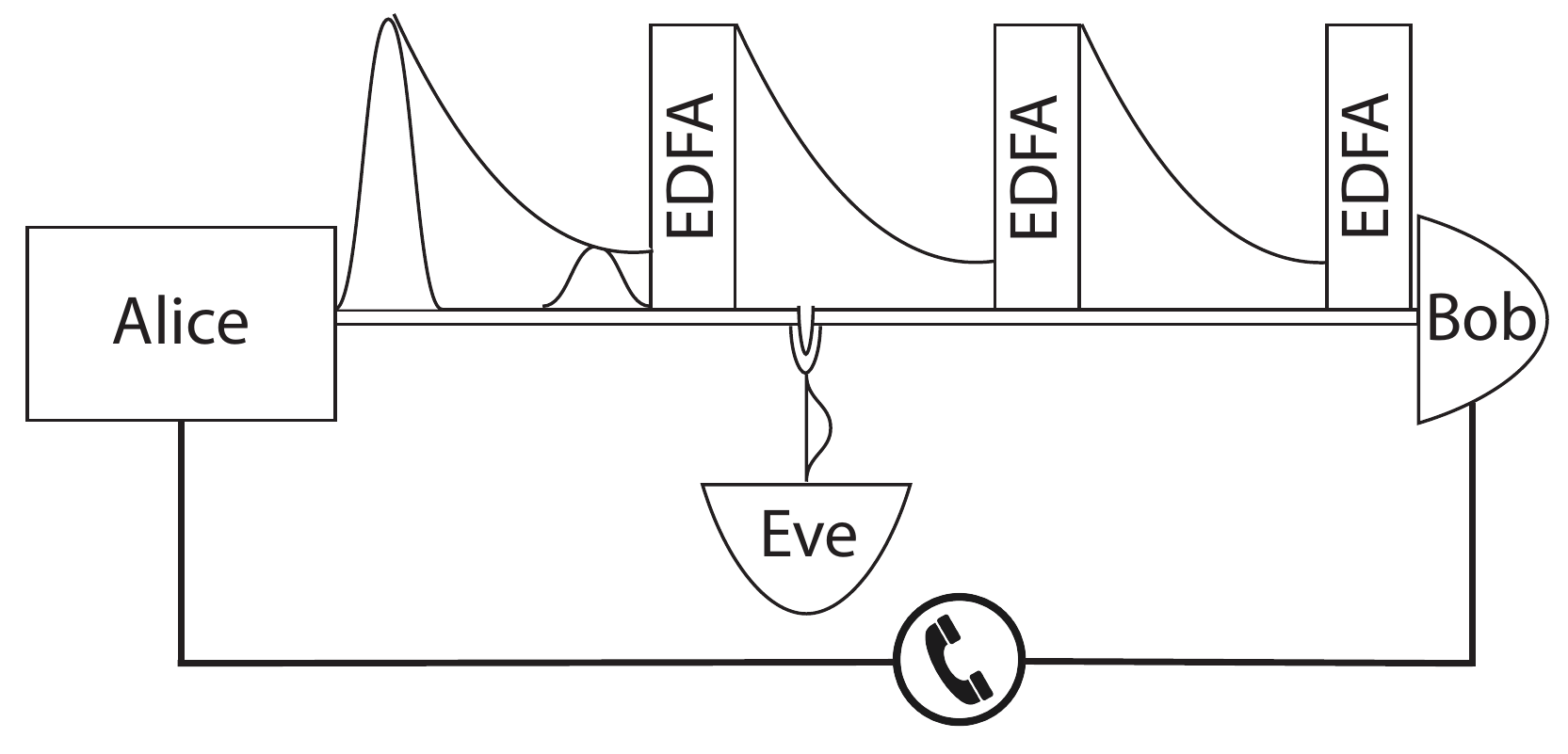}}
    \caption{
    Schematics of the setup realizing the exemplary protocol. 
    Alice encodes a random bit string into a sequence of coherent pulses and sends it to Bob via the optical fiber. 
    The pulses pass through a sequence of optical amplifiers and the resulting signals are then received and measured by Bob.
    Eve can seize part of the signal: for instance, by bending the transmitting optical fiber and detecting the transcending optical modes. 
    However, Alice and Bob monitor the losses in the line, and always know the proportion of the signal stolen by eavesdropper.
    Importantly, they can identify the exact losses caused and exploited by Eve.
    This knowledge enables Alice and Bob to adopt the most efficient bit ciphering and measurement scheme: depending on how many photons Eve intercepts, Alice picks certain values of signal intensities which are optimal as far as the informational advantage over Eve is concerned; in a concerted manner, Bob adjusts his measurement routine.
    This in particular gives the authorized parties additional leverage as far as post-selection is concerned: after transmitting and receiving the random bit string, Alice and Bob use an authenticated public classical channel to perform information reconciliation (increasing their informational advantage over Eve) and privacy amplification -- with the optimal parameters of ciphering and measurement these procedures allow to eradicate Eve's information without sacrificing too many bits. 
    } 
    \label{Fig1}
\end{figure}

\section{Method description}

We propose a method for long-distance QKD based on the signal amplification and physical control of the transmission line (see the schematics in Fig.\ref{Fig1}).
Two principal ideas behind the method are: (i) the random bits are encoded into non-orthogonal coherent pulses which are amplified by a cascade of the in-line optical amplifiers, e.g., Erbium Doped Fiber Amplifiers (EDFAs) to achieve long-distance transmission, and (ii) Alice and Bob can determine the exact proportion of the signal stolen by the eavesdropper (Eve) and distinguish it from the natural losses in the line (caused primarily by the Rayleigh scattering).
Alice and Bob use the knowledge about losses to accurately estimate their informational advantage over Eve which in turn allows them to pick the pulses' intensities, adopt the measurement routine and perform post-selection in the most efficient way, leaving Eve no information about the final shared key.

\begin{enumerate}
    \item [0.] \textit{Initial preparation.--} The reflectometry methods allows to determine losses with high precision and distinguish local losses (which could be caused by Eve) from the intrinsic natural losses homogeneous across the whole line and caused mainly by the Rayleigh and Raman scattering.
    As a part of the initial equipment setting, Alice and Bob determine the natural losses $r_0$ in the transmission line which cannot be caused by Eve.
    Bob and Alice share the value of $r_0$ via the authenticated classical communication channel.
    
    \item Alice and Bob determine a total signal loss $r_t$ in the communication channel via transmitting test pulses.
    After that, they obtain the signal loss $r_E$ caused by an eavesdropper by comparing the intrinsic signal loss $r_0$ with the total signal loss $r_t$.
    Bob and Alice share the value of $r_E$ via the authenticated classical communication channel.
    
    For example, let us consider a section of the communication channel which does not comprise amplifiers.
    If the intrinsic loss in this section is $r_0$  and Eve seizes a proportion $r_E$ of the signal, then the total loss $r_t$ is determined from the equation $(1-r_t)=(1-r_E)(1-r_0)$. 
    Further below we describe a beam splitter attack of Eve 5 at a single location along the communication channel.
    The proposed method and our analysis presented further below also generalizes to the case where Eve intrudes the communication channel at several locations.
    
    \item Using a physical random number generator (possibly quantum), Alice generates a bit sequence of length $L$.
    
    \item Alice ciphers her bit sequence into a series of $L$ coherent light pulses which she sends to Bob.
    The bits 0 and 1 are defined by the coherent states $\ket{\gamma_0}=\ket{\gamma}$ and $\ket{\gamma_1}=\ket{-\gamma}$ respectively,  -- without loss of generality, assume that $\gamma\in\mathds{R}$ -- and the value of $\gamma$ is chosen optimally given the known specific value of $r_E$.
    This means that Alice uses such coherent states that they correspond to the maximum key generation speed with respect to the losses in the channel.
    The intensity of the pulse $\ket{\pm\gamma}$ is determined by the average photon number 
    \begin{equation}
        \langle n \rangle =|\gamma|^2.
    \end{equation}

    \item The signal is amplified by a cascade of optical amplifiers installed equidistantly along the whole optical line. 
    Each amplifier compensates the losses in such a way that the amplified signal intensity equals to the initial one.
    As a coherent pulse passes through amplifiers its state becomes mixed.
    Bob receives the signal and performs the homodyne measurement, the parameters of which are again determined by the known value of $r_E$.

    \item Alice and Bob apply information reconciliation.
    Some of Bob's measurements will have inconclusive results and the corresponding bits must be discarded.
    To do so, Bob announces positions of invalid bits to Alice publicly via an authenticated public classical channel.
    
    \item Alice and Bob estimate the error rate and perform the error correction procedure.
    
    \item Alice and Bob perform privacy amplification.
    Using a special protocol, Alice and Bob produce a shorter key, which Eve has no (or negligibly small) information about.
    Once again, Alice and Bob may need to use their authenticated public classical channel.
    
    \item Alice and Bob perform steps 1 to 7 until the length of the shared key is appropriate.
\end{enumerate}

The outlined protocol is based on phase ciphering.
However, in the case of very long transmission distances, the preservation of the pulses' polarization and the efficient phase reference transmission necessary for homodyne detection may be challenging.
These practical difficulties can be lifted, by, for example, using intensity ciphering, in which case the bit values are encoded into the two pulses with different intensities.
This has already been mentioned in the Introduction and is analyzed below in Section 3.1.
%
In the main part of the paper, we will concentrate primarily on the phase ciphering approach which ensures faster information transmission. In the forthcoming publication we will analyse in detail the special measures necessary for compensating phase fluctuations hence improving considerably   the phase ciphering protocol.

\section{Signal amplification}
The crucial component of the proposed scheme is the cascade of amplifiers preserving the signal intensity necessary for achieving long distance transmission. 
We start this section with estimates illustrating how amplification impacts the signal fluctuations and the precision in detecting leakage.
We then introduce the formal framework based on the P-function representation and show how the signal state evolves under the ideal amplification process.
After that, we consider the practical case of amplification in doped fibers with the associated losses in the channel.
We further show that a cascade of amplifiers can be theoretically reduced to one effective amplifier -- we will use this formal property in further sections for the analysis of legitimate users' informational advantage over the eavesdropper.

\subsection{Preliminary estimates}

The challenge to meet is reaching the global distances of the transmission exceeding 20,000\,km. 
In order to estimate possible enhancement of the transmission distance we use, for illustrative purposes, simple and transparent estimates based on the robustness of the strength of our pulses.
Let us estimate the strength of our pulses that allows for their stable analogue amplifying, transmitting the signal without distorting its shape and phase with the minimal generated noise and, at the same time, preserving the degree of protection against eavesdropping.
A crucial component needed to ensure such an enhancement are just the standard telecom amplifiers preserving the analog signal including erbium-doped fiber amplifier and Raman amplifier. 
To give an idea how it works, we present simple estimates leaving the detailed description for the technical part of the paper.

Let us consider the control precision for the representative line of 20,000 km.
In general, the exact optimal distance between amplifiers is to be calculated. Here for the estimate we take the standard telecom practice distance of $d=50$\,km. 
On this distance the signal drops by factor of 10 (the transmission probability $T=0.1$); correspondingly the amplifying coefficient is to be $G=10$. 
Suppose that the initial test signal carries $n^{A}_{T}=10^{14}$ photons, which drops down to $T\cdot n^{A}_{T} = 0.1\cdot 10^{14} = 10^{13}$ on the 50\,km end. 
An amplifier restores it back to $G\cdot T\cdot n^{A}_{T} = 10\cdot 0.1\cdot 10^{14} = 10^{14}$ but adds noise. 
Since photons follow the Poisson statistics, the fluctuations before the amplifier are $\sqrt{T\cdot n^{A}_{T}} \simeq 3\cdot 10^6$. 
These fluctuations are amplified with the factor $G$ as well, giving 
\begin{equation}
    \delta n^G_T\simeq G \sqrt{T n^{A}_{T}}\simeq 3 \cdot 10^7.
\end{equation}
Coming through the sequence of $M$ amplifiers which add fluctuations independently, the total fluctuation raises by factor $\sqrt{M}$, giving for 400 amplifiers on the 20,000\,km line the twenty-fold fluctuation increase. 
The fluctuation on Bob's end is thus
\begin{equation}
    \delta n^{B}_{T}\simeq\sqrt{M}\,\delta n^G_T\simeq6\cdot 10^8.
\end{equation} 
At the end of the day, we have
\begin{equation}
    \delta n^{B}_{T}/n^{B}_{T}\simeq6\cdot 10^{-6}.
\end{equation}
This determines the minimum detectable leakage $r_E^\text{min}\simeq 6\cdot 10^{-6}$.
Our qualitative considerations agree with the rigorous calculations provided in Sec. 3.5.

For illustrative purposes, let us use the intensity ciphering instead of the phase ciphering.
If Eve gets $\sim 1$ photons (stealing signal in the vicinity of Alice), unity relative error would make it impossible for her to distinguish between the different pulses provided that their intensities are of the same order.
A pulse hard for eavesdropping must thus contain $n\sim 1/r_E^\text{min} \sim 10^5$ photons; in this case, the fluctuation at Bob's amounts to 
\begin{gather}
    \delta n = \sqrt{G M n} \sim 10^4;\\  
    \delta n/n\sim  0.1.
\end{gather}
This allows, especially taking into account additional informational postprocessing measures exercised by both Bob and Alice, more elaborate encoding and measuring procedures, the readable Alice to Bob transmission that cannot be intercepted by Eve. The fact that scalability behaves as square root of the line length, enables us to execute a major breakthrough in safe communication sending the decipherable signal over a global distance of 40,000\,km. 

\subsection{P-function and its evolution under amplification}

Let us introduce our theoretical framework for the rigorous description of losses and amplifications.
Consider a single photonic mode with bosonic operators $\hat a$ and $\hat a^\dag$ acting in the Fock space.
To understand the effect of the amplification on the bosonic mode state, it is most convenient to use the P-function representation of the latter. 
Such representation allows to express any density operator as a quasi-mixture of coherent states:
\be
\hat{\rho} = \int d^2 \alpha \, \t{P}(\alpha) \prjct{\alpha},
\label{P-representation}
\ee
where $d^2 \alpha\equiv d\text{Re}(\alpha)\,d\text{Im}(\alpha)$ and the quasi-probability distribution $\t{P}(\alpha)$ is not necessarily positive.
For a given state described by density matrix $\hat\rho$ the P-function can be written as 
\be
\t{P}(\alpha) = \tr \hat{\rho} :\delta(\hat a-\alpha):
\ee
where
\be
:\delta(\hat a-\alpha): = \frac{1}{\pi^2}\int d^2\beta\,  e^{\alpha \beta^* - \alpha^* \beta} e^{\beta \hat a^\dag} e^{-\beta^* \hat a},
\ee
see\,\cite{Vogel} for details.\\ 

Phase-amplification is described by a quantum channel given by
\be\begin{split}
    \t{Amp}_{G=\cosh^2(g)}: \hat{\rho} \mapsto \t{Amp}_{G}[\hat{\rho}]&=\tr_b \hat{U}_g\,   \hat{\rho} \otimes \prjct{0}_b\, \hat{U}_g^\dag\\
    \hat{U}_g &= e^{g (\hat{a}^\dag \hat{b}^\dag - \hat{a} \, \hat{b})},
\end{split}
\ee
where $g$ is the interaction parameter characterizing the amplifier, $G = \cosh^2(g)$ is the factor by which the intensity of the input signal is amplified (as we will explicitly see in the following formulae), and annihilation operator $\hat b$ corresponds to the auxiliary mode starting in the vacuum states.
An explicit Kraus representation of the channel can be written as
\be
\t{Amp}_{G}(\hat{\rho}) =\sum_{n=0}^\infty \hat{K}_n \hat{\rho} \hat{K}_n
\ee
with $\hat{K}_n = \frac{\tanh^{n}(g)}{\sqrt{n!}} \hat{a}^{\dag n} \cosh(g)^{-\hat{a}\, \hat{a}^\dag}$ see e.g.\,\cite{Sekatski}.

To show how the P-function of a state changes under the amplification process let us consider a simple situation where the input signal is in the pure coherent state $\prjct{\gamma}$ with the corresponding initial P-function $\t{P}_i(\alpha)= \delta(\alpha-\gamma)$ (delta-function on complex numbers).
After the amplification the P-function becomes
\be
P(\alpha;\gamma,g) = \tr  :\delta(\hat a-\alpha): \t{Amp}_{G}\left[\prjct{\gamma} \right].
\ee
Bearing in mind that
\be\label{eq: canonical amp}
\t{Amp}^*_{G}[\hat{a}]=\hat{U}_g^\dag\,  \hat{a} \, \hat{U}_g = \cosh(g) \hat{a} + \sinh(g) \hat{b}^\dag,
\ee
it is easy to see that
\be\boxed{
P(\alpha; \gamma, g) = \frac{1}{\pi (G-1)} \exp\left(-\frac{|\alpha- \sqrt{G} \gamma|^2}{G-1}\right).}
\label{P-function}
\ee
In other words, the output state is a mixture of normally distributed states centered around $\ket{\sqrt{G} \gamma}$; the width of the distribution is $(G-1)/\sqrt{2}$.

\subsection{Amplification in doped fibers and losses}

In Er/Yt doped fiber the photonic mode propagates through the inverted atomic medium. 
To keep the medium inverted, a seed laser of a different frequency co-propagates with the signal photonic mode in the fiber and is then filtered out at the output by means of wavelength-division multiplexing (WDM).
The interaction between the inverted atoms at position $z$ and propagating light field mode $\hat{a}$ is precisely given by the Hamiltonian
\be
H = \ii (\hat{a}^\dag \hat{b}_z^\dag - \hat{a} \hat{b}_z),
\ee
where $\hat b_z$ corresponds to a collective decay  of one of the atoms at $z$. 
Hence, the evolution of the signal mode after its propagation through EDFA is set by a composition of infinitesimal phase-amplifications which, as we show in the next subsection, can be effectively reduced to a single amplification channel.

In practice, the performance of EDFA suffers from technical limitations, which come in addition to the amplification limits on added quantum noise. 
These limitations are mainly caused by two factors: (i) the atomic population may be not completely inverted throughout the media, (ii) coupling imperfection between the optical mode and EDFA or optical fiber.
Both of these mechanisms can be taken into account as a loss channel acting on the state before the amplification, as shown in\,\cite{Sanguinetti}.

Let us introduce the loss channel describing all possible losses in the line.
Equation~\eqref{eq: canonical amp} describes the action of amplifier on the annihilation operator in the Heisenberg picture.
In the same way we can express the canonical transformation associated with losses
\be\begin{split}
    \t{Loss}_T^*[\hat{a}] = \hat{\bar U}_\lambda^\dag \hat{a} \hat{\bar U}_\lambda &= \cos(\lambda) \hat{a} + \sin(\lambda) \hat{c} \\ & = \sqrt{T} \hat{a} + \sqrt{1-T} \hat{c},\\
    T& =\cos^2(\lambda),
\end{split}
\ee
where $\lambda$ is the interaction parameter, $T$ is the proportion of the transmitted signal, the annihilation operator $\hat{c}$ corresponds to the initially empty mode which the lost photons go to, and $\hat{\bar U}_\lambda=e^{\lambda \hat{a}^\dag \hat{c} - \lambda \hat{a} \, \hat{c}^\dag}$.\\

\subsection{Composition of amplifiers and losses}
\begin{figure}[t]
    \noindent\centering{
    \includegraphics[width=0.8\columnwidth]{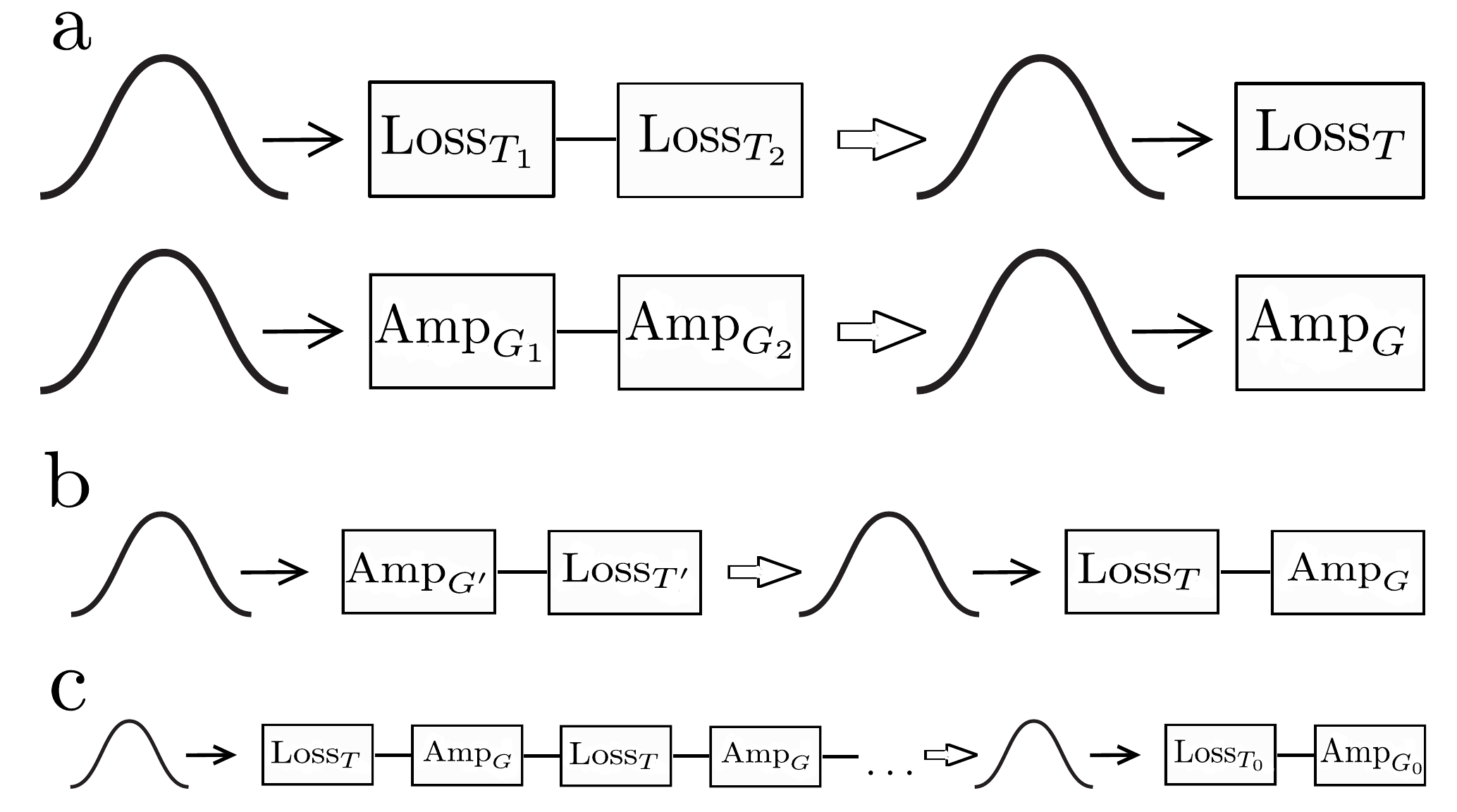}}
    \caption{
    (a) Two loss or amplification channels can be reduced to one. (b) Loss and amplification channels can be effectively rearranged. (c) A series of losses and amplifiers can be reduced to one pair of loss and amplification.
    } 
    \label{Fig2}
\end{figure}

In our cryptographic scheme, the amplification is used to recover the optical signal after it suffers from losses. 
Long-distance QKD requires a cascade of amplifiers, in which case signal's evolution is determined by a sequence of multiple loss and amplification channels.
In this section we prove that any such sequence can be mathematically reduced to a composition of one loss and one amplification channels.
We will later adopt this simple representation for the informational analysis of our protocol.

\subsubsection*{Statement 1. Two loss or amplification channels can be reduced to one}

First, we show that a pair of loss or amplification channels can be effectively reduced to the one channel, see Fig.\ref{Fig2}(a).
To that end, let us consider two consequent loss channels:
\be\begin{split}
    (\t{Loss}_{T_2}\circ \t{Loss}_{T_1})^* [\hat{a}] &= \hat{\bar U}_{\lambda_1}^\dag \hat{\bar U}_{\lambda_2}^\dag \hat{a} \, \hat{\bar U}_{\lambda_2} \hat{\bar U}_{\lambda_1}\\
    &=\t{Loss}_{T_1}(\sqrt{T_2}\hat{a} + \sqrt{1- T_2} \hat{c}_2)\\
    & = \sqrt{T_2 T_1} \hat{a} + \sqrt{T_2(1-T_1)}\hat{c}_1 + \sqrt{1- T_2} \hat{c}_2\\
    & = \sqrt{T_2 T_1} \hat{a} + \sqrt{1-T_1T_2} \hat{c},
\end{split}
\ee
where we defined operator 
\be
\hat{c} =\frac{\sqrt{T_2(1-T_1)}\hat{c}_1 + \sqrt{1- T_2} \hat{c}_2)}{\sqrt{1-T_1T_2}},
\ee
acting on the vacuum state and satisfying the canonical commutation relation $[\hat{c},\hat{c}^\dag]=1$. 
We can thus represent two channels in a form of one effective channel:
\be\begin{split}
    \t{Loss}_{T_2}\circ \t{Loss}_{T_1} = \t{Loss}_{(T=T_1T_2)}
\end{split}
\ee
The same reasoning applies to amplifiers:
\be
    \t{Amp}_{G_2}\circ \t{Amp}_{G_1} = \t{Amp}_{(G=G_2G_1)}
\ee\\

\subsubsection*{Statement 2. Loss and amplification channels can be effectively rearranged}

Let us show that a composition of an amplification channel followed by a loss channel can be mathematically replaced with a pair of certain loss and amplification channels acting in the opposite order, see Fig.\ref{Fig2}(b).
Consider the transformation corresponding to the amplification followed by the loss
\be\label{eq: AT}\begin{split}
    (\t{Loss}_{T'}\circ \t{Amp}_{G'})^*[\hat{a}] &=   \hat{U}_{g'}^\dag \hat{\bar U}_{\lambda'}^\dag \hat{a} \, \hat{\bar U}_\lambda \hat{U}_g \\
    & = \t{Amp}_{G'}^* [\sqrt{T'} \hat{a} + \sqrt{1-T'} \hat{c}]\\
    & =\sqrt{T' G'} \hat{a} +\sqrt{(1-T')} \hat{c} + \sqrt{T'(G'-1)} \hat{b}^\dag.
\end{split}
\ee
In the case of the opposite order we have
\be\label{eq: TA}\begin{split}
    (\t{Amp}_G\circ\t{Loss}_T)^*[\hat{a}] &= \hat{\bar U}_\lambda^\dag \hat{U}_g^\dag \hat{a} \, \hat{U}_g \hat{\bar U}_\lambda\\
    & = \t{Loss}_T^* [\sqrt{G} \hat{a} + \sqrt{G-1} \hat{b}^\dag]\\
    & =\sqrt{T G} \hat{a} +\sqrt{G(1-T)} \hat{c} + \sqrt{G-1} \hat{b}^\dag.
\end{split}
\ee
It is easy to see that the two transformations are identical if 
\be\label{permute}\boxed{\begin{aligned}
    \t{Loss}_{T'}\circ \t{Amp}_{G'} &= \t{Amp}_G\circ\t{Loss}_T\\
    T &= \frac{G' T'}{(G'-1)T'+ 1},\\
    G &= (G'-1)T '+1.
\end{aligned}}
\ee
In other words, the two types of channels "commute" provided that the parameters are modified in accord with these relation.
In particular, the parameters in the equation above are always physically meaningful $G\geq 1, 0\leq T \leq 1$, meaning that we can always represent loss and amplification in form of a composition where loss is followed by amplification (the converse is not true).\\

\subsubsection*{Statement 3. A series of losses and amplifiers can be reduced to one pair of loss and amplification}

Let us finally show that a sequence of loss and amplification channels can be mathematically represented as one pair of loss and amplification, see Fig.\ref{Fig2}(c).
Consider the transformation
\be
\Phi_M = (\t{Amp}_G\circ \t{Loss}_T)^{\circ M},
\ee
corresponding to the series of $M$ identical loss and amplification channels, for which we want to find a simple representation. 
According to Statement 2, we can effectively move all losses to the right end of the composition, i.e., permute the channels in such a way that all the losses act before amplification.
Every time a loss channel with transmission probability $T_{(i)}$ is moved before an amplifier with amplification factor $G_{(i)}$, the parameters are transformed in accord with Eq.\,(\ref{permute}):
\be\begin{split}
T_{(i)} &\mapsto T_{(i+1)}=\frac{G_{(i)} T_{(i)}}{(G_{(i)}-1) T_{(i)} + 1},\\
G_{(i)} &\mapsto G_{(i+1)}=(G_{(i)}-1)T_{(i)} +1. 
\end{split}
\ee
In our sequence we can pairwise transpose all neighbouring loss with amplifier (starting with the first amplifier and the second loss).
After repeating this operation $M-1$ times, bearing in mind Statement 1, we find that
\be\begin{split}
\Phi_M&= \t{Amp}_{G_{(0)}}\circ \t{Amp}_{G_{(1)}}\circ \dots \circ\t{Amp}_{G_{(M-1)}}\\
        &\quad \circ \t{Loss}_{T_{(M-1)}}\circ \t{Loss}_{T_{(M-2)}}\circ \dots \circ \t{Loss}_{T_{(0)}}\\
        & = \t{Amp}_{G_\circ} \circ \t{Loss}_{T_\circ},
\end{split}
\ee
where 
\be
\begin{split}
    T_\circ &= \prod_{i=0}^{M-1} T_{(i)},\\
    G_\circ &= \prod_{i=0}^{M-1} G_{(i)},
\end{split}
\ee
i.e., a series of losses and amplifiers is equivalent to the loss channel of transmission $T_\circ$ followed by the amplifier with amplification factor $G_\circ$.

Note now that the value $\mu\equiv G_{(i)}T_{(i)}=G T$ cannot be changed by permutations. 
Let us define
\begin{equation}
    F_{(i)} = (G_{(i)}-1) T_{(i)}+1,
\end{equation}
and bear in mind that 
\be\begin{split}
    F_{(i+1)} = (G_{(i+1)}-1) T_{(i+1)}+1
    = \frac{(F_{(i)}-1)}{F_{(i)}} TG + 1
    =  \mu \left(\frac{F_{(i)}-1}{F_{(i)}}\right) + 1.
\end{split}
\ee\\
We can write
\be\begin{split}
T_{(i+1)} &= \frac{ T G}{F_{(i)}},\\
G_{(i+1)} &= F_{(i)},
\end{split}
\ee
and
\be
\begin{split}
    G_\circ &=  G \prod_{i=0}^{M-2}F_{(i)},\\
    T_\circ &=  \frac{T (TG)^{M-1}}{ \prod_{i=0}^{M-2} F_{(i)}}= \frac{(T G)^M}{G_\circ}.
\end{split}
\ee
Let us find the explicit form of $G_\circ$ and $T_\circ$ by solving the recurrence relation.
Define $A_{n}$ and $B_{n}$ through the relation
\be
F_{(n-1)} = \frac{A_{n}}{B_{n}}.
\label{def Fn}
\ee
Then
\be
F_{(n+1)} = \frac{(\mu+1) F_{(n)}-\mu}{F_{(n)}} = \frac{(\mu+1) A_{n+1}-\mu B_{n+1} }{A_{n+1}}.
\label{relation for Fn through An and Bn}
\ee
It follows from \eqref{def Fn} and \eqref{relation for Fn through An and Bn}  that $B_{n+1}= A_n $ and
\be
A_{n+1} = (\mu+1) A_n-\mu B_n = (\mu+1) A_n-\mu A_{n-1}.
\ee
We see that the solution of this equation has a form
\be
A_{n} = c_1 + c_2 \mu^n,
\ee
where $c_1$ and $c_2$ are the constants, which are determined by $F_0 =(G-1)T+1$: we take $A_1 = (G-1)T+1$ and $A_0 =1$, and obtain
\be
c_1 =\frac{T-1}{G T-1} \qquad c_2 = \frac{(G-1) T}{G T-1}.
\ee
Notably, the product $\Pi_{n=0}^{M-2} F_{(n)}$ appearing in the final expression becomes relatively simple
\be\begin{split}
    \prod_{n=0}^{M-2} F_{(n)} &= \prod_{n=0}^{M-2} \frac{A_{n+1}}{A_n} = \frac{A_{M-1}}{A_0}\\
    &= \frac{(G-1) (G T)^M+G (T-1)}{G (G T-1)},
\end{split}
\ee
and we have
\be\boxed{\begin{aligned}
\Phi_M &= (\t{Amp}_G\circ \t{Loss}_T)^{\circ M}= \t{Amp}_{G_\circ} \circ \t{Loss}_{T_\circ},\\
    G_\circ &= \frac{(G-1) (G T)^M+G (T-1)}{G T-1},\\
    T_\circ &= \frac{(TG)^M}{G_\circ}.
\end{aligned}}
\ee

The case of $TG=1$ is particularly interesting as the average intensity of the transmitted signal remains preserved (which is different from the total output intensity as it has the noise contribution). 
In the limit $G\to 1/T$ we have
\be\boxed{\begin{aligned}
\label{equivGT}
    G_\circ &= G(M(1-T)+T),\\
    T_\circ&=\frac{T}{M (1-T)+T}. 
\end{aligned}}
\ee

\subsection{Fluctuations}

Let us calculate the fluctuation of the number of photons in a pulse after it passes through a sequence of $M$ loss regions and amplifiers.
Let $|\gamma|^2$ be the input intensity;
as follows from\,\eqref{P-representation} and\,\eqref{P-function}, the average number of photons $n$ in the output signal is
\begin{equation}
\begin{split}
    n = \langle\hat{a}^\dagger\hat{a}\rangle = |\gamma|^2 + G_\circ - 1,
\end{split}
\end{equation}
where $G_\circ$ and $T_\circ$ are given by Eq.\,\eqref{equivGT}. 
The variance of the output photon number is
\begin{align}
    \delta n^2 &= \langle(\hat{a}^\dagger\hat{a})^2\rangle - (\langle\hat{a}^\dagger\hat{a}\rangle)^2 = G M (1 - T)(G M (1 - T) + 1) + |\gamma|^2(2 G M (1 - T) + 1).
    \label{fluctuation of intensity after M amp-s}
\end{align}
In the limit $|\gamma|^2\gg GM\gg1$ we obtain the same result as we got from the qualitative considerations in Sec. 3.1: 
\begin{equation}
\label{fluct}
    \delta n\simeq \sqrt{n G M}.
\end{equation}

\section{Control of the transmission line}

To monitor the eavesdropper's activity, Alice sends, at appropriate intervals, special test pulses (individual or many, see the discussion below) and cross-checks the intensities with Bob. 
The test pulses should comprise a large number of photons, but must not, however, damage Bob's detection equipment.
By producing and analysing the corresponding scattering matrix, Alice and Bob can determine the losses in the channel.
To prevent an inconspicuous Eve's intrusion into the optical fiber, the physical loss control should be conducted constantly and should not halt even during the pauses in the key distribution.
We emphasize that the authorized parties can discriminate between the losses of the general origin and losses that are caused and can be exploited by Eve.

If the optical fiber is properly installed, it should not have points of significant inflections and crude junctions.
Then, most of the intrinsic natural losses in the line occur due to Rayleigh scattering.
Such losses are distributed across the whole line.
Therefore, Eve cannot pick up the dissipated signal effectively, unless she has an antenna covering a significant part of the line;
moreover, as we discussed in the Introduction, the effective deciphering of the dissipated signal requires a multitude of Maxwell demon-like devices, which is not practically feasible.
The only option remaining to the eavesdropper is to take away part of the signal deliberately, that is to create and exploit the losses additional to the natural ones (namely, by bending the optical fiber intentionally).
Alice and Bob can identify and measure such artificial losses.
To do that, they must first determine the magnitude of losses not associated with the eavesdropper's activity; this can be done by measuring the losses appearing homogeneously across the whole line before the beginning of the protocol. 
After that, Alice and Bob can precisely determine the newly appearing local leaks of the signal (with proportion $r_E$) possibly intercepted by Eve.
This knowledge ensures the most efficient ciphering and measurements routines, determining in turn the post-selection procedure.

In the case that Eve knows the parameters of the test pulses, she can completely seize them and send the fake ones to mask her presence.
In order to prevent that, Alice should parametrize the test sequences randomly and compare the parameters with Bob only after he measures the pulses.
This arrangement would force Eve to first measure the pulses and then reproduce them which in turn would prolong the transmission.
Such a delay can be easily detected by Alice and Bob.
Possible testing protocols:
\begin{enumerate}
    \item \textit{Individual pulses.}-- Alice sends a single testing pulse the parameters of which are chosen randomly.
    The preparation of the pulse implies generating an auxiliary random bit sequence and translating it into the random intensity, phase, length and shape of the pulse.
    After Bob measures the test pulse, he verifies its parameters with Alice, and they determine the losses in the channel.
    \item \textit{Sequence of pulses.}-- Alice sends a sequence of test pulses in which she encodes an auxiliary random sequence.
    This involves generating an auxiliary random sequence and ciphering it in a sequence of pulses.
    Bob measures the pulses, verifies the encoded message with Alice, after what they determine the losses.
\end{enumerate}

Let $\tau_S$ be the length of the signal pulse, and $\tau_T$ be the total length of a sequence of test pulses (test pulses may look completely like the signal ones, but their sequence must contain much more photons than one signal pulse).
Both types of pulses can be characterized by the same constant power, e.g., $P=20$\,mW, but $\tau_T$ must be much greater than $\tau_S$, e.g., $\tau_T=1$\,ms and $\tau_S=1$\,ns.
The average number of photons in the test pulse is $n_T^A=\frac{P\tau_T}{h\nu}\sim10^{14}$, where $\nu$ is the light frequency.
The measurement error on Bob's side is determined by Eq.\,\eqref{fluct}: $\delta n_T^B\sim 6\cdot 10^8$.
The test pulse allows to detect leakage of magnitudes
\begin{equation}
    r_E\geq\delta T_T= \delta n_T/\langle n_T \rangle= 6 \cdot 10^{-6}.
\end{equation}
Similar control and analysis of the reflected signal must be performed on Alice's end.

\section{Measurement scheme}
\begin{figure}[t]
    \noindent\centering{
    \includegraphics[width=0.8\columnwidth]{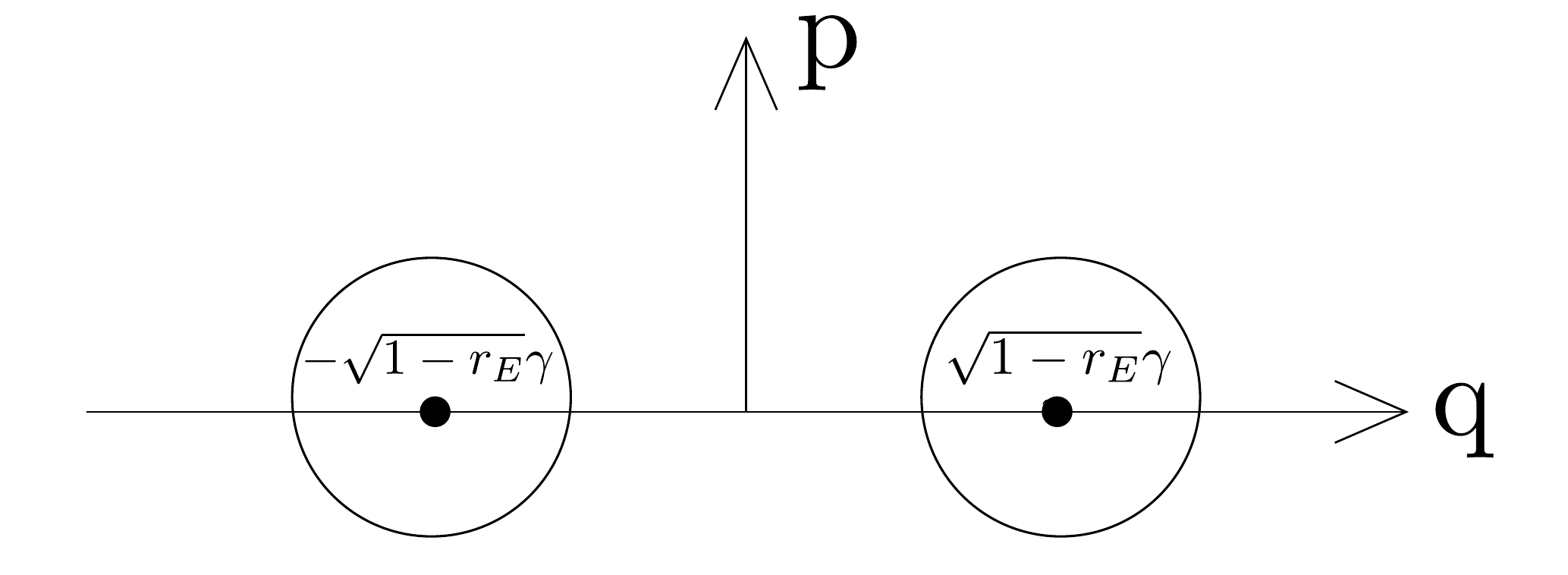}}
    \caption{
    Optical phase diagram of states corresponding to bit values 0 and 1 after they pass through the sequence of losses and amplifications.
    The two output states constitute Gaussians centered at $\pm\sqrt{1-r_E}\gamma$, where $r_E$ is the proportion of signal stolen by Eve. 
    } 
    \label{Fig3}
\end{figure}

 
The state of the optical signal can be described in terms of its quadratures given by operators
\begin{equation}
\begin{gathered}
    \hat{q} = \frac{\hat{a}^{\dagger} + \hat{a}}{2},\\
    \hat{p} = \frac{i(\hat{a}^{\dagger} - \hat{a})}{2}.
\end{gathered}
\end{equation}
These operators represent the real and imaginary parts of the signal's complex amplitude, and by measuring one of the quadratures one can distinguish between different signals.

Bob needs to distinguish between two states $\ket{\gamma_0}=\ket{\gamma}$ and $\ket{\gamma_1}=\ket{-\gamma}$ (with $\gamma\in\mathds{R}$) transformed by losses and amplifiers, -- two Gaussians with centers laying on the real axis ($q$-axis) of the optical phase space, see the phase diagram displayed in Fig.\,\ref{Fig3} -- and for this he measures the $\hat{q}$-quadrature.
Bob does his measurements by means of homodyne detection described by the following operators:
\begin{equation}
    \begin{gathered}
        \hat{E}_0 = \int\limits^{\infty}_{\Theta} d q \ket{q}\bra{q},\\
        \hat{E}_1 = \int\limits^{-\Theta}_{-\infty} d q \ket{q}\bra{q},\\
        \hat{E}_\text{fail} = \hat{I} - \hat{E}_0 - \hat{E}_1,
    \end{gathered}
\end{equation}
where $\ket{q}$ is the eigenstate of $\hat{q}$, and the parameter $\Theta$ is tuned by Bob depending on the amount of losses possibly stolen by Eve, see the discussion below. 
\begin{figure}[t]
    \noindent\centering{
    \includegraphics[width=0.8\columnwidth]{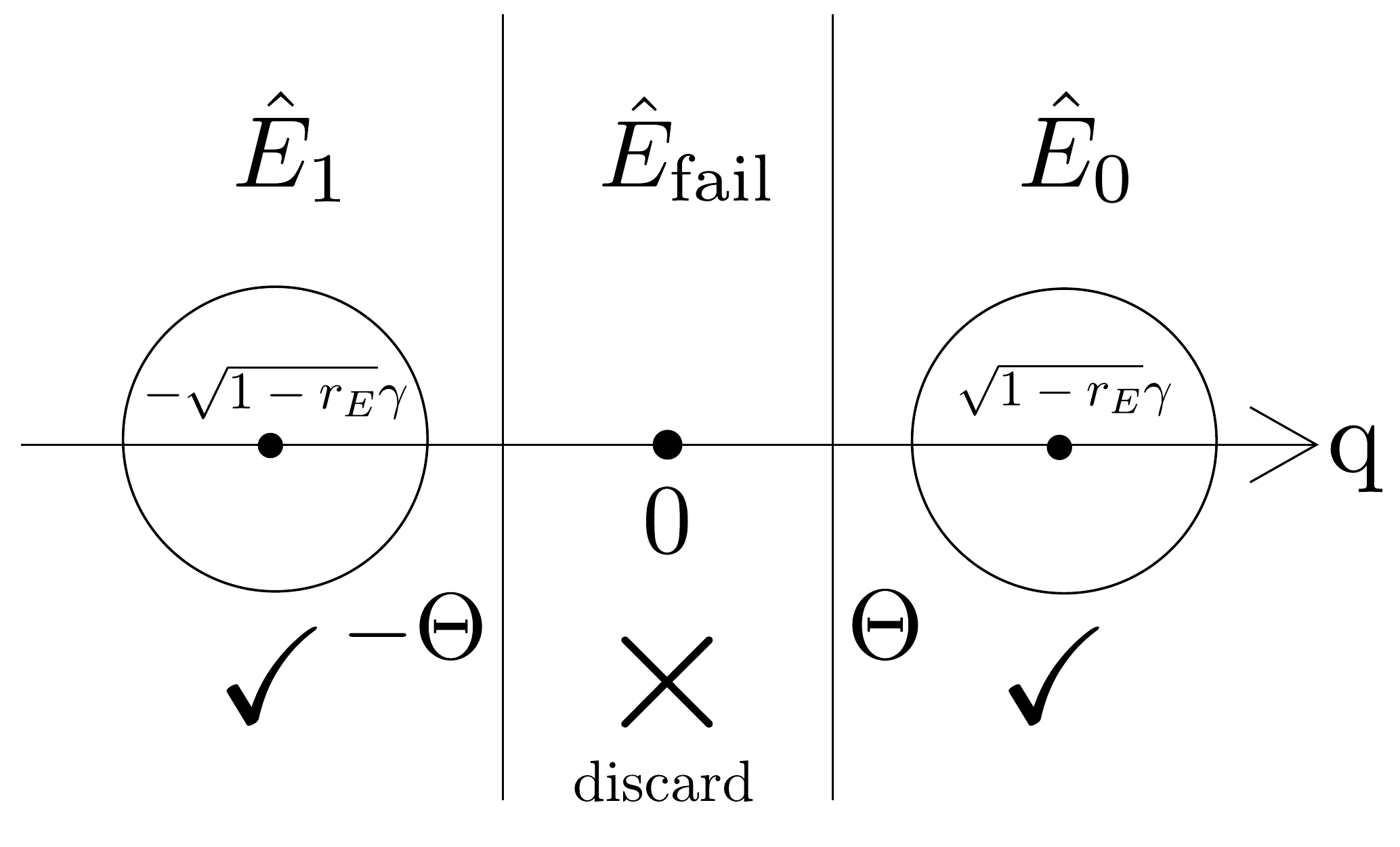}}
    \caption{
    Schematic representation of the measurement operators employed by Bob.
    The two signal states are Gaussians overlapping the most in the region of phase space corresponding to $\hat{E}_\text{fail}$.
    Therefore, Alice and Bob consider the associated outcome inconclusive and discard it on the post-selection stage. 
    The outcomes associated with $\hat{E}_0$ and $\hat{E}_1$ are considered conclusive.
    The value of $\Theta$ is varied to perform the most efficient post-selection procedure as far as the final key generation rate is concerned.
    } 
    \label{Fig4}
\end{figure}
Here, $\hat{E}_{0(1)}$ determines the bit value 0(1), whereas $\hat{E}_\text{fail}$ is associated with the bad outcome and the respective bit must be discarded by Alice and Bob on the stage of post-selection. 
By looking at Fig.\,\ref{Fig4}, one can see that $\hat{E}_\text{fail}$ corresponds to the phase space region where the two states (Gaussians) overlap the most, and thus the associated outcome is inconclusive.

\section{Error estimate and correction}

After the measurements and post-selection procedure of discarding invalid bits, Alice and Bob must perform error correction procedure.
The quadrature value $q$ obtained from homodyne measurement, allows to estimate the probability of error in the corresponding bit -- one can easily compute the conditional error probability for every $q$.
In reality, the error rate is also determined by channel imperfections and Eve's detrimental activities.
Therefore, in practice, instead of the theoretical prognosis, the error correction procedure should be mainly predicated on the direct measurement of an error, e.g., by disclosing a part of the raw key to observe the error.

One option for the practical error estimation is to disclose one half of the raw key. 
But if the raw key is long enough, already a relatively short part can provide an accurate error estimate. 
According to this method, Alice and Bob use their public authenticated channel to select a number of bit positions in the raw key and publicly announce the corresponding bit values. 
Then, using Bayes' theorem, they can make a guess about the expected error rate for the remaining part of the raw key.
Alternatively, instead of the bit values Alice and Bob can disclose the parity bits for some selected blocks of raw key positions. 
This method gives a better estimate for small error rate values, but is worse for high error rates. 
The decision about the block size can be made taking into account the theoretical estimates based on the observed values of $q$. 
For example, if the raw error estimate is approximately 6\%, blocks of length $10$ can be used, since the probability for parity bits mismatch in this case is approximately 36\%, which is large, but still below 50\%, meaning that the parity data reveal a lot of information about the real error rate. 

After estimating the error rate, Alice and Bob can carry out the error correction procedure. 
To do that, they can use the low-density parity-check (LDPC) codes\,\cite{MacKay}.
The input for such codes are the probabilities of zero or one at each bit position and the syndrome of the correct bit string -- the set of parity bits which are sufficient to correct the errors, taking into account the a priori probabilities for each position. 
LDPC codes are particularly good for error correction after homodyne measurement, since the measurement result $q$ itself allows to calculate the probabilities for correct and erroneous results.

After Alice and Bob correct errors in a (possibly small) part of the raw key, they have to take into account the number of errors in this part to yield a more accurate error estimate for the remaining key.
We propose the following adaptive procedure: Alice and Bob first take a relatively short subset of the original raw key (which size depends on the codeword's length, e.g., 1000 bits), and apply error correcting procedure which is designed for high error rates (e.g., 10\% in the case that the preliminary crude estimation gave just 5\% error probability). 
After correcting error in this small subset, Alice and Bob know the number of errors therein, and have a better error rate estimate for the remaining part of the key.
Then, they should take another short (e.g., once again, 1000 bits) subset and perform error correction according to the new improved error rate estimate, and so on. 
With each iteration, the error estimate becomes more accurate making the error correction procedure more efficient. 
This method can be applied without having initial error estimate at all which can save Alice and Bob a large part of the raw key.

The error correction procedure discloses some information about the key. For linear codes like LDPC codes, one syndrome bit discloses no more than one bit of information about the key, thus the syndrome length is appropriate upper bound for the information leakage.

\section{Privacy amplification}

Although after the error correction procedure, Alice and Bob share the same bit string, which is still correlated with Eve, thus it cannot be used as the final secret key.
The privacy amplification procedure\,\cite{Bennett_PA, Bennett_PAgeneralized, Brassard_PA}, aimed at eradicating Eve's information, produces a new, shorter, bit string.
This new string can finally be used as a secret key as Eve does not possess any (or almost any) information about it.

To eliminate the eavesdropper information, Alice and Bob can, for instance, use universal hashing method\,\cite{Carter_universal_hashing}.
This method requires them to initially agree on the family $H$ of hash functions $h \in H$.
At the privacy amplification stage, they randomly select such a function $h:\{0,1\}^{l_1}\rightarrow \{0,1\}^{l_2}$ from this family, that it maps the original bit string of length $l_1$ to the final key of length $l_2$. 
If Eve is estimated to have $e$ bits of information about the raw key, $l_2$ must be equal to $l_1-e$.

One example of $H$ is the Toeplitz matrices family\,\cite{krawczyk_Toeplitz}. 
Alice and Bob can use a random binary Toeplitz matrix $T$ with $l_1$ rows and $l_2$ columns. 
Then they represent their bit string as a binary vector $v$, and the final key $k$ is given by
$$k = T\cdot v.$$

\section{Eve's attack}
\begin{figure}[t]
    \noindent\centering{
    \includegraphics[width=0.8\columnwidth]{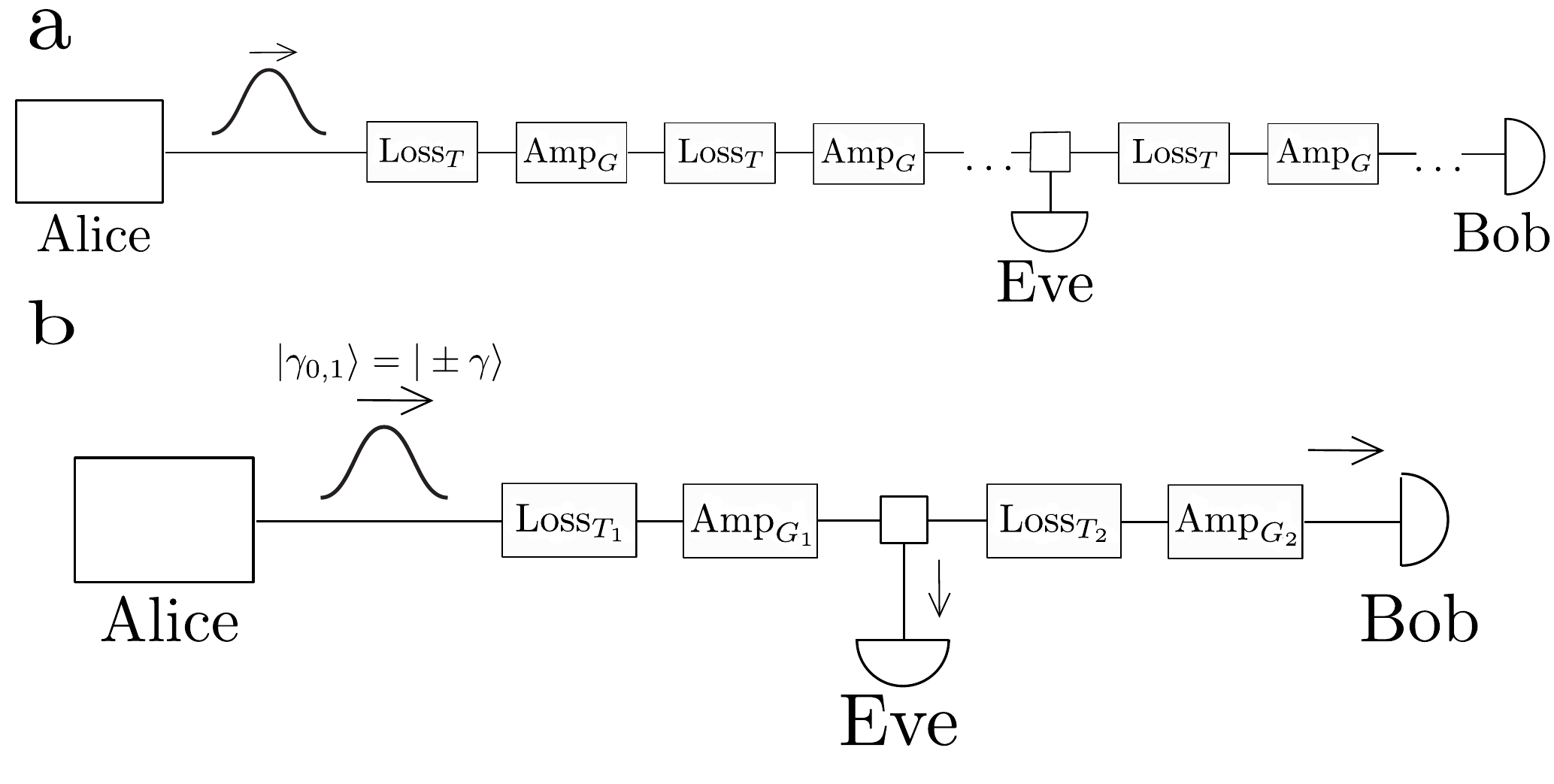}}
    \caption{
    (a) Beam splitter attack on the protocol. 
    Eve seizes part of the signal somewhere along the optical line.
    (b) Scheme equivalent to (a). 
    The losses and amplifiers before and after the point of Eve's intervention are represented by two pairs of loss and amplification channels defined by the parameters $\{T_1,\,G_1\}$ and $\{T_2,\,G_2\}$ respectively.
    } 
    \label{Fig5}
\end{figure}
In this section, we demonstrate how the protocol works in the case where Eve performs the beam splitter attack seizing the part of the signal somewhere along the optical line as shown in Fig.\,\ref{Fig5}(a) (we will use the term "beam splitter" in referring to the point of Eve's intervention into the line).
The beam splitter is ideal, meaning that there is no reflection in Alice's direction. 
If the signal intensity incident to the beam splitter is 1, then intensity $r_E$ goes to Eve, and $1-r_E$ goes to Bob's direction.

Quantum cryptography also studies attacks exercising the partial blocking of the signal and the subsequent unauthorised substitution of the blocked part.
However, any attack like this will inevitably and permanently (even if Eve at some point decides to disconnect from the line) affect the scattering matrix of the transmission line, and hence will be detected by the legitimate users.  
Therefore, we do not concentrate on this kind of attacks here.
Furthermore, at this point we accept that the attacks specifically focusing on amplifiers can be reduced to the beam splitter kind of attacks. The associated subtleties and details will be the subject of our forthcoming publication.

\subsection{Losses and amplifiers}
The proportion of the transmitted signal on the distance $d$ between two neighbouring amplifiers is determined by
\begin{equation}
    T = 10^{-\mu d},
\end{equation}
where $\mu=1/50\,\text{km}^{-1}$ is the parameter of losses typical for the optical fibers.
As was mentioned before, the amplification factor of each amplifier is $G=1/T$.
Let $D_{AB(AE)}$ be the distance between Alice and Bob (Alice and Eve), then the numbers of amplifiers before and after the beam splitter $M_1$ and $M_2$ are given by
\begin{gather}
    M_1 = D_{AE}/d,\\
    M_2 = (D_{AB}-D_{AE})/d.
\end{gather}
As we showed previously, the scheme can be simplified by reducing the losses and amplifications before and after the beam splitter to two loss and aplification pairs with the parameters $\{T_1,\,G_1\}$ and $\{T_2,\,G_2\}$ respectively
\be
\begin{split}
    T_1 = \frac{T}{M_1(1-T)+T} = \frac{10^{-\mu d}}{(1-10^{-\mu d})D_{AE}/d+10^{-\mu d}}, G_1 = \frac{1}{T_1},\\
    T_2 = \frac{T}{M_2(1-T)+T} = \frac{10^{-\mu d}}{(1-10^{-\mu d})(D_{AB}-D_{AE})/d+10^{-\mu d}}, G_2 = \frac{1}{T_2}.
\end{split}
\ee

\subsection{Evolution of systems' state}
Let us describe the progressive evolution of the combined systems' state.
The initial state of Alice's random bit ($A$) -- her random number generator -- and the corresponding signal ($S$) is given by
\begin{equation}
    \hat{\rho}^\text{i}_{AS}= \frac{1}{2}\ket{0}\bra{0}_A \otimes \ket{\gamma}\bra{\gamma}_S +
    \frac{1}{2}\ket{1}\bra{1}_A \otimes \ket{-\gamma}\bra{-\gamma}_S.
\end{equation}
As the signal undergoes transformations associated with losses and amplifications, the state of the $AS$-system just before the signal passes the beam splitter is given by
\begin{equation}
\label{AS_BS}
\begin{gathered}
    \hat{\rho}^{\rightarrow \Box}_{AS} =
     \frac{1}{2}\ket{0}\bra{0}_A \otimes \int d^2 \alpha P_{\sqrt{T_1}\gamma, G_1}(\alpha)\ket{\alpha}\bra{\alpha}_S\\ +
    \frac{1}{2}\ket{1}\bra{1}_A \otimes \int d^2 \alpha P_{-\sqrt{T_1}\gamma, G_1}(\alpha)\ket{\alpha}\bra{\alpha}_S.
\end{gathered}
\end{equation}
Defining 
\begin{equation}
\begin{gathered}
    \hat{\rho}^{\rightarrow \Box}_{AS}[\alpha] = 
    \frac{1}{2} \ket{0}\bra{0}_A \otimes \ket{\alpha}\bra{\alpha}_S + \frac{1}{2} \ket{1}\bra{1}_A \otimes \ket{-\alpha}\bra{-\alpha}_S,
\end{gathered}
\end{equation}
we can rewrite Eq.\,(\ref{AS_BS}) as
\begin{equation}
    \hat{\rho}^{\rightarrow \Box}_{AS} = \int d^2 \alpha P_{\sqrt{T_1}\gamma, G_1}(\alpha)\hat{\rho}^{\rightarrow \Box}_{AS}[\alpha].
\end{equation}

Just after the signal passes the beam splitter, the state of the joint system comprising Alice's random bit ($A$), the signal going to Bob ($S$) and signal seized by Eve ($E$) is described by
\begin{equation}
    \hat{\rho}^{\Box \rightarrow}_{ASE} = \int d^2 \alpha P_{\sqrt{T_1}\gamma, G_1}(\alpha)\hat{\rho}^{\Box \rightarrow}_{ASE}[\alpha] 
\end{equation}
where
\begin{equation}
\begin{gathered}
    \hat{\rho}^{\Box \rightarrow}_{ASE}[\alpha] = \frac{1}{2}\ket{0}\bra{0}_A \otimes \ket{\sqrt{1 - r_E}\alpha}\bra{\sqrt{1 - r_E}\alpha}_S \otimes\ket{\sqrt{r_E}\alpha}\bra{\sqrt{r_E}\alpha}_E\\
    + \frac{1}{2}\ket{1}\bra{1}_A \otimes \ket{-\sqrt{1 - r_E}\alpha}\bra{-\sqrt{1 - r_E}\alpha}_S \otimes\ket{-\sqrt{r_E}\alpha}\bra{-\sqrt{r_E}\alpha}_E,
\end{gathered}
\end{equation}
and $r_E$ is the proportion of signal stolen by Eve.
After the signal undergoes the second sequence of losses and amplifiers and just before it is measured by Bob, the state of the joint system is
\begin{equation}
    \hat{\rho}^{\rightarrow \text{Bob}}_{ASE} = \int d^2 \alpha P_{\sqrt{T_1}\gamma, G_1}(\alpha)\hat{\rho}^{\rightarrow \text{Bob}}_{ASE}[\alpha] 
\end{equation}
with
\begin{equation}
\begin{gathered}
    \hat{\rho}^{\rightarrow \text{Bob}}_{ASE}[\alpha] = \frac{1}{2}\sum_{a=0,1} \ket{a}\bra{a}_A \otimes \left(\int d^2 \beta P_{\sqrt{(1 - r_E)T_2}(-1)^a \alpha, G_2}(\beta)\ket{\beta}\bra{\beta}_S\right) \otimes\ket{(-1)^a\sqrt{r_E}\alpha}\bra{(-1)^a\sqrt{r_E}\alpha}_E.
\end{gathered}
\end{equation}

Bob receives the signal, measures it and, together with Alice, performs post-selection, which lays in discarding the bits associated with the fail-outcome by communicating through the classical channel.
The probability that Bob's measurement outcome is $b=\{0,\,1\}$ given that Alice sent bit is $a=\{0,\,1\}$ can be written as
\begin{equation}
\label{prob1}
    p(b|a)=\tr_{ASE}\left[{\left(2\cdot\ket{a}\bra{a}_A \otimes \hat{E}_b\otimes \mathds{1}_E\right) \hat{\rho}^{\rightarrow \text{Bob}}_{ASE}}\right]=\int d^2 \alpha P_{\sqrt{T_1}\gamma, G_1}(\alpha) p_{(-1)^a\alpha}(b|a),
\end{equation}
where 
\begin{equation}
\label{prob2}
\begin{gathered}
    p_{(-1)^a \alpha}(b | a) = \int d^2 \beta P_{\sqrt{(1 - r_E)T_2}(-1)^a\alpha, G_2}(\beta)\bra{\beta}\hat{E}_b\ket{\beta}_S.
\end{gathered}
\end{equation}
Thus, the probability of a conclusive outcome which means that the bit will not be discarded on the stage of post-selection is 
\begin{equation}
\label{checkmark}
    p(\checkmark) =\frac{1}{2}\sum_{a,b=0,1} p(b|a) = \int d^2 \alpha P_{\sqrt{T_1}\gamma, G_1}(\alpha)\sum\limits_{b = 0, 1}\frac{p_{\alpha}(b | 0) + p_{-\alpha}(b | 1)}{2}.
\end{equation}
The final state of Alice's random bit ($A$), Bob's memory device storing the measurement outcome ($B$) and the signal stolen by Eve ($E$) after the post-selection, i.e., conditional to the successful measurement outcome, is 
\begin{equation}
    \hat{\rho}^{\text{f}}_{ABE} = \int d^2 \alpha P_{\sqrt{T_1}\gamma, G_1}(\alpha)\hat{\rho}^{\text{f}}_{ABE}[\alpha] 
    \label{final}
\end{equation}
where
\begin{equation}
\begin{gathered}
    \hat{\rho}^{\text{f}}_{ABE}[\alpha] = \frac{1}{2}\ket{0}\bra{0}_{A} \otimes \left(\sum\limits_{b = 0, 1} \frac{p_{\alpha}(b | 0)}{p(\checkmark)}\ket{b}\bra{b}_{B}\right) \otimes \ket{\sqrt{r_E}\alpha}\bra{\sqrt{r_E}\alpha}_E\\
    + \frac{1}{2}\ket{1}\bra{1}_{A} \otimes \left(\sum\limits_{b = 0, 1} \frac{p_{-\alpha}(b | 1)}{p(\checkmark)}\ket{b}\bra{b}_{B}\right) \otimes \ket{-\sqrt{r_E}\alpha}\bra{-\sqrt{r_E}\alpha}_E.
    \label{rho_final}
\end{gathered}
\end{equation}

\subsection{Probabilities}

To obtain the probabilities $p(b|a)$, $a,b\in\{0,\,1\}$, we must first calculate $|\langle\beta|q\rangle|^2$:
\begin{equation}
    |\langle\beta|q\rangle|^2 = \bra{\beta} \left( \frac{1}{2\pi}\int d p e^{i p(\hat{q} - q)}\right) \ket{\beta}= \frac{1}{2\pi}\int d p \bra{\beta}e^{i p\left(\frac{\hat{a}^{\dagger} + \hat{a}}{2} - q\right)}\ket{\beta}
    = \sqrt{\frac{2}{\pi}}e^{-2(\text{Re}\beta - q)^2}.
\end{equation}
After substituting $|\langle\beta|q\rangle|^2$ into Eqs.\,(\ref{prob1}, \ref{prob2}) we obtain 
\begin{equation}
\label{p00}
    \begin{gathered}
    p(0|0) =
    \frac{1}{2}\left\lbrace 1 - \text{erf}\left(\frac{\sqrt{2}\left[\Theta - \gamma\sqrt{G_1 T_1 G_2 T_2(1 - r_E)}\right]}{\sqrt{1+ 2 (G_2 - 1) + 2 G_2 T_2(1 - r_E)(G_1 - 1)}}\right)\right\rbrace,
    \end{gathered}
\end{equation}
\begin{equation}
    \begin{gathered}
    p(1| 0) =
    \frac{1}{2}\left\lbrace 1 - \text{erf}\left(\frac{\sqrt{2}\left[\Theta + \gamma\sqrt{G_1 T_1 G_2 T_2(1 - r_E)}\right]}{\sqrt{1+ 2 (G_2 - 1) + 2 G_2 T_2(1 - r_E)(G_1 - 1)}}\right)\right\rbrace,
    \end{gathered}
\end{equation}
\begin{equation}
    \begin{gathered}
    p( 0 | 1) =
    \frac{1}{2}\left\lbrace 1 - \text{erf}\left(\frac{\sqrt{2}\left[\Theta + \gamma\sqrt{G_1 T_1 G_2 T_2(1 - r_E)}\right]}{\sqrt{1+ 2 (G_2 - 1) + 2 G_2 T_2(1 - r_E)(G_1 - 1)}}\right)\right\rbrace,
    \end{gathered}
\end{equation}
\begin{equation}
\label{p11}
    \begin{gathered}
    p( 1 | 1) =
    \frac{1}{2}\left\lbrace 1 - \text{erf}\left(\frac{\sqrt{2}\left[\Theta - \gamma\sqrt{G_1 T_1 G_2 T_2(1 - r_E)}\right]}{\sqrt{1+ 2 (G_2 - 1) + 2 G_2 T_2(1 - r_E)(G_1 - 1)}}\right)\right\rbrace,
    \end{gathered}
\end{equation}
where
\begin{equation}
    \text{erf}(x) = \frac{2}{\sqrt{\pi}}\int\limits^{x}_{0} e^{-t^2} d t .
    \label{erf}
\end{equation}

\subsection{Eve's information}
Hereafter we will use the concepts of the quantum von Neumann entropy and conditional entropy.
For quantum system $X$ with the density matrix $\hat\rho_X$ the entropy is defined as
\begin{equation}
    H_{\hat\rho_X}(X)=\tr\left[\hat\rho_X \log \hat\rho_X\right],
\end{equation}
where $\log\equiv\log_2$.
For the pair of quantum systems $X$ and $Y$, with the states of $X$-system and the joint system described by the density matrices $\hat\rho_X$ and $\hat\rho_{XY}$, respectively, the conditional entropy can be written as
\begin{equation}
    H_{\hat\rho_{XY}}(Y|X)=H_{\hat\rho_{XY}}(XY)-H_{\hat\rho_{X}}(X).
\end{equation}

Let us estimate Eve's information about the raw key (per 1 bit) after post-selection but before the error correction stage
\begin{equation}
\label{iae}
    I(A,\,E) = H_{\hat{\rho}^{\text{f}}_{A}}(A)-H_{\hat{\rho}^{\text{f}}_{AE}}(A|E) = 1 - H_{\hat{\rho}^{\text{f}}_{AE}}(A|E).
\end{equation}
Conditional entropy $H_{\hat{\rho}^{\text{f}}_{AE}}(A|E)$ is determined by the final density matrix of $AE$-system which given Eq.\,(\ref{final}) and the fact that 
\begin{equation}
\begin{gathered}
   p_{\alpha}(0 | 0) + p_{\alpha}(1 | 0) = p_{-\alpha}(0 | 1)+p_{-\alpha}(1 | 1),
\end{gathered}
\end{equation}
can be written as
\begin{equation}
    \hat{\rho}^{\text{f}}_{AE} =\tr_B\hat{\rho}^{\text{f}}_{ABE} =  \int d^2 \alpha Q_{\checkmark}[\alpha]\hat{\rho}^{\text{f}}_{AE}[\alpha] 
\end{equation}
with
\begin{equation}
\begin{gathered}
    \hat{\rho}^{\text{f}}_{AE}[\alpha] = \frac{1}{2}\ket{0}\bra{0}_{A} \otimes\ket{\sqrt{r_E}\alpha}\bra{\sqrt{r_E}\alpha}_E + \frac{1}{2}\ket{1}\bra{1}_{A} \otimes\ket{-\sqrt{r_E}\alpha}\bra{-\sqrt{r_E}\alpha}_E,\\
    Q_{\checkmark}[\alpha] = \frac{[p_{\alpha}(0 | 0) + p_{\alpha}(1 | 0)]P_{\sqrt{T_1}\gamma, G_1}(\alpha)}{p(\checkmark)}.
\end{gathered}
\end{equation}

To find the lower bound of Eve's entropy (and therefore estimate the maximum of her information about the key) we consider a situation where Eve has some auxiliary register of variable $\alpha$ ($R_E$) and introduce the joint $AER_E$ state:
\be
\hat{\rho}^\text{f}_{AER_E} = \frac{1}{2} \int d^2\, \alpha Q_\checkmark[\alpha] \sum_{a=0,1} \prjct{a}_A\otimes \prjct{(-1)^a\sqrt{r_E}\alpha}_E\otimes \prjct{\t{reg}(\alpha) }_{R_E}.
\ee
Here the register's states satisfy $\langle\t{reg}(\alpha)|\t{reg}(\alpha')\rangle=\delta^{(2)}(\alpha-\alpha')$. 
We note that first, by tracing out the register we recover the original state of $AE$-system
\be
\tr_{R_E} \hat{\rho}^\text{f}_{AER_E} = \hat{\rho}^\text{f}_{AE},
\ee
and second, the monotonicity of conditional entropy implies
\be
H_{\hat{\rho}^\text{f}_{AE}}(A|E)\geq H_{\hat{\rho}^\text{f}_{AER_E}}(A|E R_E)= H_{\hat{\rho}^\text{f}_{AER_E}}(AER_E) - H_{\hat{\rho}^\text{f}_{ER_E}}(ER_E).
\ee
The latter inequality simply states that after discarding the register, Eve can only lose information about the sent bit. 
To this end, $H_{\hat{\rho}^\text{f}_{AER_E}}(A|E R_E)$ is the lower bound of $H_{\hat{\rho}^\text{f}_{AE}}(A|E)$.

Matrix $\hat{\rho}^\text{f}_{AER_E}$ can be rewritten as
\be
\hat{\rho}^\text{f}_{AER_E} = \sum_i \int d^2 \alpha \, Q_\checkmark[\alpha] \lambda_i^{(\alpha)} \ \prjct{\psi_i^{(\alpha)}} \otimes \prjct{\t{reg}(\alpha)}_{R_E},
\ee
where $\lambda_i^{(\alpha)}$ and $\ket{\psi_i^{(\alpha)}}$ are the eigenvalues and eigenstates of $\hat{\rho}^\text{f}_{AE}[\alpha]$. 
We thus yield
\be
\begin{split}
H_{\hat{\rho}^\text{f}_{AER_E}}(AER_E) &= - \sum_i \int d^2 \alpha Q_\checkmark[\alpha] \lambda_i^{(\alpha)} \log(Q_\checkmark[\alpha] \lambda_i^{(\alpha)})+C\\
& =- \sum_i \int d^2 \alpha Q_\checkmark[\alpha] \lambda_i^{(\alpha)} (\log(Q_\checkmark[\alpha]) + \log( \lambda_i^{(\alpha)})) +C\\
& =  - \int d^2 \alpha Q_\checkmark[\alpha] \log(Q_\checkmark[\alpha]) 
- \int d^2 \alpha Q_\checkmark[\alpha] \sum_i \lambda_i^{(\alpha)} \log(\lambda_i^{(\alpha)}) +C\\
& = - \int d^2 \alpha Q_\checkmark[\alpha] \log(Q_\checkmark[\alpha]) 
+ \int d^2 \alpha Q_\checkmark[\alpha] H_{\hat{\rho}^{\text{f}}_{AE}[\alpha]}(AE) +C,
\end{split}
\ee
where $C$ is an additional correcting term, which we do not need to calculate explicitly.
In the same way we obtain 
\be
H_{\hat{\rho}^\text{f}_{ER_E}}(ER_E) = - \int d^2 \alpha Q_\checkmark[\alpha] \log(Q_\checkmark[\alpha]) 
+ \int d^2 \alpha Q_\checkmark[\alpha] H_{\hat{\rho}^\text{f}_{E}[\alpha]}(E)+C.
\ee
Combining the two expressions we obtain 
\be\begin{split}
     H_{\hat{\rho}^\text{f}_{AE}}(A|E)\geq H_{\hat{\rho}^\text{f}_{AER_E}}(A|E R_E) &= H_{\hat{\rho}^\text{f}_{AER_E}}(AER_E) - H_{\hat{\rho}^\text{f}_{ER_E}}(ER_E) \\
     &= \int d^2 \alpha Q_\checkmark[\alpha] (H_{\hat{\rho}^\text{f}_{AE}[\alpha]}(AE)- H_{\hat{\rho}^\text{f}_{E}[\alpha]}(E))\\
     &= \int d^2 \alpha Q_\checkmark[\alpha] H_{\hat{\rho}^\text{f}_{AE}[\alpha]}(A|E)\\
     &=1 - \int d^2 \alpha Q_{\checkmark}[\alpha] h\left(\frac{1 - |\langle\sqrt{r_E}\alpha|-\sqrt{r_E}\alpha\rangle|}{2}\right)\\
     &= 1- \int d^2 \alpha \,Q_\checkmark[\alpha] \, h\left(\frac{1+e^{-2\, r_E |\alpha|^2}}{2}\right).
\end{split}
\ee
where
\begin{equation}
    h(p) = -p\log(p) - (1 - p)\log(1 - p)
\end{equation}
is the binary entropy.
Now, we can use the Jensen's inequality 
\be
\mean{h(x)} \leq h(\mean{x}) \implies 1-\mean{h(x)} \geq 1- h(\mean{x}),
\ee
where $\mean{x}\equiv \int d^2 \alpha \,Q_\checkmark[\alpha] \,x$:
\be
\label{hae}
H_{\hat{\rho}^\text{f}_{AE}}(A|E) \geq 1- h\left( \frac{1+\mean{e^{-2\, r_E |\alpha|^2}}}{2}\right)
\ee
where 
$\mean{e^{-2\, r_E |\alpha|^2}} =  \int d^2 \alpha \,Q_\checkmark[\alpha] \,e^{-2\, r_E |\alpha|^2}$.
We find
\begin{equation}
\label{mean}
\begin{gathered}
\mean{e^{-2\, r_E |\alpha|^2}} =
\frac{e^{-\frac{2 r_E G_1 T_1|\gamma|^2}{1 + 2 r_E \epsilon_1}}}{p(\checkmark)[1 + 2 r_E \epsilon_1]}\Bigg\{1 - \frac{1}{2}\text{erf}\Bigg(\frac{\sqrt{2(1 + 2 r_E \epsilon_1)}\Big[\Theta - \frac{\sqrt{(1 - r_E)G_2 T_2 G_1 T_1}\gamma}{1 + 2 r_E \epsilon_1}\Big]}{\sqrt{2 \epsilon_2 + 1 + 2 (1 - r_E)G_2 T_2\epsilon_1 + 2 r_E(2 \epsilon_2 + 1)\epsilon_1}}\Bigg)\\
- \frac{1}{2}\text{erf}\Bigg(\frac{\sqrt{2(1 + 2 r_E \epsilon_1)}\Big[\Theta + \frac{\sqrt{(1 - r_E)G_2 T_2 G_1 T_1}\gamma}{1 + 2 r_E \epsilon_1}\Big]}{\sqrt{2 \epsilon_2 + 1 + 2(1 - r_E)G_2 T_2\epsilon_1 + 2 r_E(2 \epsilon_2 + 1)\epsilon_1}}\Bigg)\Bigg\},
\end{gathered}
\end{equation}
where $\epsilon_{1,2}\equiv G_{1,2}-1$.

By substituting Eq.\,(\ref{mean}) into Eq.\,(\ref{hae}) and using Eq.\,(\ref{iae}) we obtain the upper bound for Eve's information.

\subsection{Bob's error rate}
Let us estimate Bob's error rate.
Bob's information per 1 bit about the key after post-selection but before error correction is given by
\begin{equation}
    I(A,\,B) = H_{\hat{\rho}^{\text{f}}_{A}}(A)-H_{\hat{\rho}^{\text{f}}_{AB}}(A|B) = 1 - H_{\hat{\rho}^{\text{f}}_{AB}}(A|B)
\end{equation}
should ideally be equal to 1.
Bob's error rate is therefore determined by the conditional entropy $H_{\hat{\rho}^{\text{f}}_{AB}}(A|B)$:
\begin{equation}
     H_{\hat{\rho}^{\text{f}}_{AB}}(A | B) = H_{\hat{\rho}^{\text{f}}_{AB}}(AB) - H_{\hat{\rho}^{\text{f}}_{B}}(B).
\end{equation}
It follows from Eq.\,\eqref{final} that 
\begin{equation}
    H_{\hat{\rho}^{\text{f}}_{AB}}(AB) = -\sum\limits_{a, b = 0, 1}\frac{p(b | a)}{2 p(\checkmark)}\log\left(\frac{p(b | a)}{2 p(\checkmark)}\right),
\end{equation}
\begin{equation}
\begin{split}
    H_{\hat{\rho}^{\text{f}}_{B}}(B) = -\sum\limits_{b = 0, 1}\frac{p(b | a = 0) + p(b | a = 1)}{2 p(\checkmark)}\log\left(\frac{p(b | a = 0) + p(b | a = 1)}{2 p(\checkmark)}\right),
\end{split}
\end{equation}
and the probabilities $p(b|a)$ and $p(\checkmark)$ are given by Eqs.\,(\ref{p00}-\ref{p11}) and (\ref{checkmark}). Notice, that post-selection is symmetric
\begin{equation}
    p(0 | 0) + p(1 | 0) = p(0 | 1) + p(1 | 1) = p(\checkmark).
\end{equation}
Then
\begin{equation}
    H_{\hat{\rho}^{\text{f}}_{AB}}(A|B) =  h\left(\frac{p(0 | 0)}{p(\checkmark)}\right).
    \label{I_AB}
\end{equation}

After the error correction procedure, Bob's information about the key becomes $\tilde I(A,B)=1$, but Eve's information increases and one can estimate it as 
\begin{equation}
    \tilde I(A,E) = I(A,E)+H_{\hat{\rho}^{\text{f}}_{AB}}(A|B) \leq 1-\min H_{\hat{\rho}^\text{f}_{AE}}(A | E) + H_{\hat{\rho}^{\text{f}}_{AB}}(A|B).
\end{equation}

\begin{figure}[t]
    \noindent\centering{
    \includegraphics[scale=0.80]{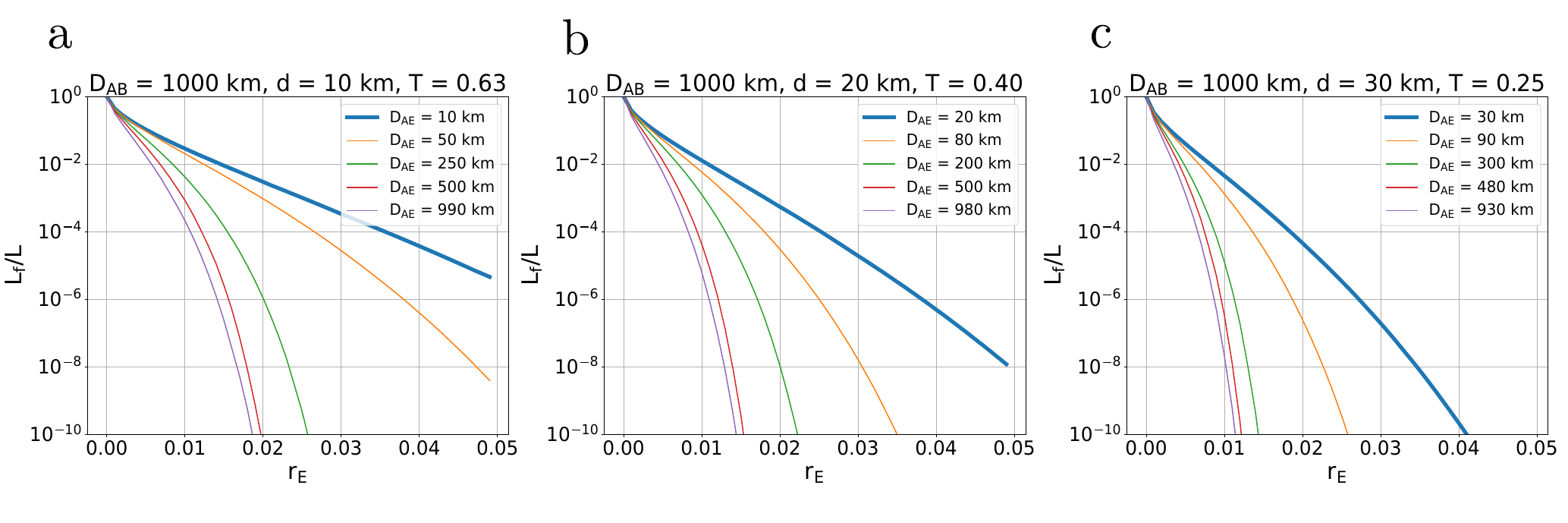}
    \caption{$L_f/L$ as function of $r_E$ for different values of distance $d$ between two neighbouring amplifiers: (a) $d=10$\,km; (b) $d=20$\,km; (c) $d=30$\,km; $D_{AB}=1000$\,km is fixed. 
    The optimal parameters: $\gamma\sim\theta\sim10$ for $r_E\sim 0.01$, and $\gamma\sim\theta\sim10^4$ for $r_E = 0.00$.
    We see that the final rate depends on Eve's position in the line: the closer she is to Bob (small $D_{AE}$), the more information she knows about the bit sequence, due to the correlations imposed by the amplifiers.
    }
    \label{Fig6}}
\end{figure}
\section{Key rate}
Now, we can estimate the length $L_f$ of the final key after post-selection, error correction and privacy amplification:
\begin{equation}
\label{speed}
    L_f= p(\checkmark)L\cdot\left(1 - \tilde I(A,E)\right)=p(\checkmark)L\cdot\left(\min H_{\hat{\rho}^\text{f}_{AE}}(A | E)- H_{\hat{\rho}^\text{f}_{AB}}(A | B)\right),
\end{equation}
where $L$ is the length of the random bit string, originally generated by Alice.
If $L$ is understood as the number of bits generated in a unit of time, then Eq.\,\eqref{speed} reflects the final key generation rate.
This equation in its explicit form, which is too cumbersome to be written here, includes two parameters -- the amplitude of signal $\gamma$ and the measurement parameter $\Theta$ -- which Alice and Bob can vary depending on $r_E$ to ensure the best rate (e.g., by numerically maximizing the function in Eq.\,\eqref{speed}).
Furthermore, the equation also includes the distance $d$ between two neighboring amplifiers and the distances $D_{AB}$ and $D_{AE}$ between the participants of the action -- these are parameters which we consider fixed.

\begin{figure}[t]
    \noindent\centering{
    \includegraphics[scale=0.80]{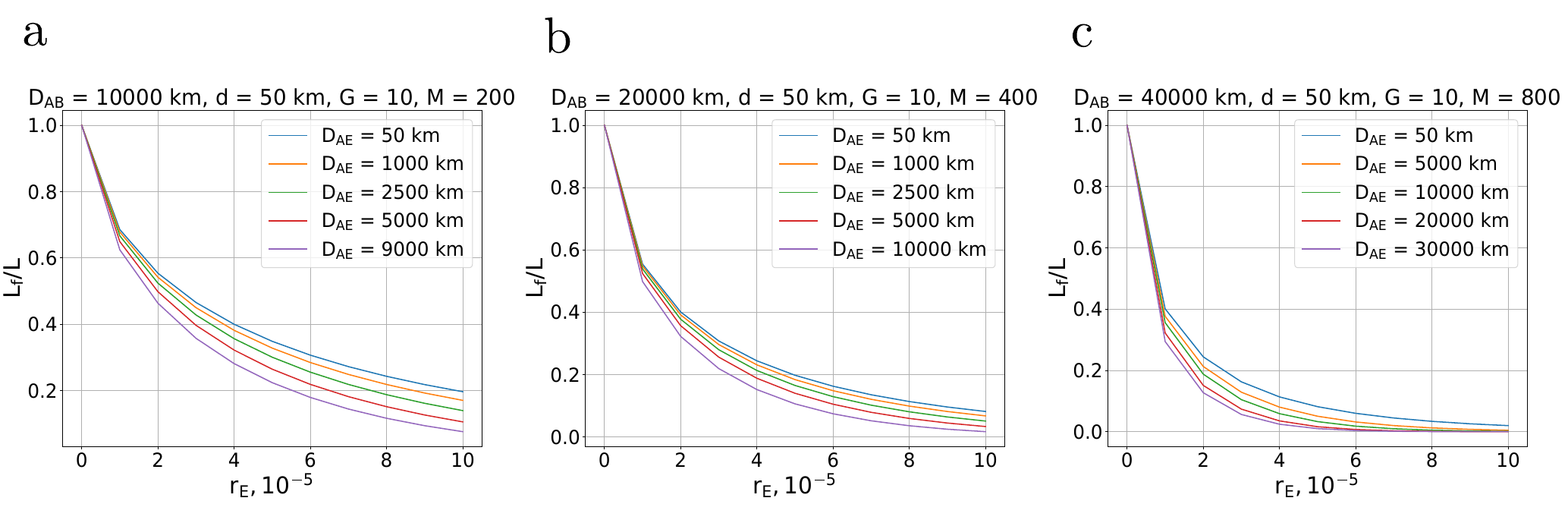}
    \caption{$L_f/L$ as function of $r_E$ for different values of distance $D_{AB}$ between Alice and Bob:
    (a) $D_{AB}=10,000$\,km; (b) $D_{AB}=20,000$\,km; (c) $D_{AB}=40,000$\,km;
    $d=50$\,km is fixed. 
    The scales of the plot imply the theoretically allowed loss detection precision.
    The optimal parameters: $\gamma\sim\theta\sim10^2$ for $r_E\sim 10^{-5}$, and $\gamma\sim\theta\sim10^4$ for $r_E = 0$.}
    \label{Fig7}}
\end{figure}

Figure\,\ref{Fig6} plots $L_f/L$ as function of $r_E$ for different values of distance $d$ between two neighbouring amplifiers. 
For every value of $r_E$ the parameters $\gamma$ and $\Theta$ are such that they maximize $L_f/L$.
Since there are correlations imposed by the amplifiers, for Eve it is beneficial to intercept signal near Bob.
The secure and fast communication can be established if $L_f/L\gtrsim10^{-6}$ (provided that the initial random number generation rate $L\sim 1\,$Gbit/s), thus Alice and Bob can allow Eve to steal no more than few percent.

Figure\,\ref{Fig7} shows $L_f/L$ as function of $r_E$ for large $D_{AB}$ in the assumption that we can technically measure $r_E$ with theoretically allowed precision.
We see that no matter from what point in the line Eve steals the signal, the precision detection of local losses allow for the secure communication over global distances $\geq20,000$\,km.


\bibliographystyle{unsrturl}
\bibliography{bib}

\end{document}